\documentclass[5p,times]{elsarticle}

\usepackage{lineno,hyperref}
\usepackage{subcaption}
\usepackage{xcolor}
\usepackage{tikz}
\usepackage{amsmath,amssymb,amsfonts}
\usepackage[capitalize]{cleveref}
\usetikzlibrary{calc,positioning,trees,shapes,matrix}

\modulolinenumbers[5]

\date{4 August 2021}

\journal{Parallel Computing}

\begin{document}

\newcommand{\iparagraph}[1]{~\\\noindent{\em #1.}}

\begin{frontmatter}

\title{Measurement and Analysis of GPU-accelerated Applications with HPCToolkit\footnotemark[1]}

\author{Keren~Zhou}
\author{Laksono~Adhianto}
\author{Jonathon~Anderson}
\author{Aaron~Cherian}
\author{Dejan~Grubisic}
\author{Mark~Krentel}
\author{Yumeng~Liu}
\author{Xiaozhu~Meng}
\author{John~Mellor-Crummey}
\address{Department of Computer Science, Rice University, Houston, TX}

\begin{abstract}
To address the challenge of performance analysis on the US DOE's forthcoming exascale supercomputers,  Rice University has been extending its HPCToolkit
performance tools to support measurement and analysis of GPU-accelerated applications. To help developers understand the
performance of accelerated applications as a whole, HPCToolkit's
measurement and analysis tools attribute metrics to calling contexts
that span both CPUs and GPUs. To measure GPU-accelerated applications
efficiently, HPCToolkit employs a novel wait-free data structure to
coordinate monitoring and attribution of GPU performance. To
help developers understand the performance of complex GPU code
generated from high-level programming models, HPCToolkit constructs sophisticated approximations of call path profiles for GPU
computations. To support fine-grained analysis and tuning, HPCToolkit
uses PC sampling and instrumentation to measure and attribute GPU
performance metrics to source lines, loops, and inlined code. 
To supplement fine-grained measurements, HPCToolkit can measure GPU kernel executions using hardware performance counters. 
To provide a view of how an execution evolves over time, HPCToolkit
can collect, analyze, and visualize call path traces within and across nodes. 
Finally, on NVIDIA GPUs, HPCToolkit can derive and attribute a
collection of useful performance metrics based on measurements using
GPU PC samples.
We illustrate HPCToolkit's new
capabilities for analyzing GPU-accelerated applications with several
codes developed as part of the Exascale Computing Project.
\end{abstract}


\end{frontmatter}


\footnotetext[1]{The initial version of this article was submitted to Parallel Computing on 30 October 2020. This is the accepted version of the article submitted on 4 August 2021. Please cite this article as K. Zhou, L. Adhianto, J. Anderson et al., Measurement and analysis of GPU-accelerated applications with HPCToolkit, Parallel Computing (2021), DOI: \url{https://doi.org/10.1016/j.parco.2021.102837}.
This work is licensed under the Creative Commons Attribution-NonCommercial-NoDerivs 2.0 Generic License. To view a copy of this license, visit http://creativecommons.org/licenses/by-nc-nd/2.0/ or send a letter to Creative Commons, PO Box 1866, Mountain View, CA 94042, USA.}

\section{Introduction}
In recent years, compute nodes accelerated with Graphics Processing Units (GPUs) have become increasingly common in supercomputers.
In June 2020, six of the world's ten most powerful supercomputers employ GPUs~\cite{top500}.
Each of the US DOE's forthcoming exascale systems Aurora, 
Frontier,
and El Capitan
are based on GPU-accelerated compute nodes.

While GPUs can provide high performance, 
without careful design GPU-accelerated applications may underutilize
GPU resources by 
idling compute units, employing insufficient thread parallelism, or exhibiting poor data locality.
Moreover, while higher-level programming models such as RAJA~\cite{hornung2014raja}, Kokkos~\cite{edwards2014kokkos}, OpenMP~\cite{bertolli2014coordinating}, and DPC++~\cite{dpc++} can simplify development of HPC applications, they can increase the difficulty of tuning GPU kernels (routines compiled for offloading to a GPU) for high performance by separating developers from many key details, such as what GPU code is generated and how it will be executed.

To harness the full  power of GPU-accelerated nodes, application developers need tools to identify performance problems.
Performance tools for GPU-accelerated programs employ \textit{trace} and \textit{profile} views.
A trace view presents events that happen over time on each process, thread, and GPU stream. 
A profile view aggregates performance metrics over the time dimension. 
Most  performance tools that support GPUs~\cite{knupfer2008vampir, shende2006tau, january2015allinea, schulz2008open, crayreveal, reinders2005vtune, nsightsystem, nsightcompute, nvprof} only provide trace and profile views with the name of each GPU kernel.
For large-scale GPU-accelerated applications, it is often difficult to understand how performance problems arise without associating
the cost of GPU kernels with their CPU calling contexts.
Manually associating the performance of GPU kernels with their CPU calling contexts is difficult when
a kernel is called from many contexts or when the name of a kernel is the result of a C++ template instantiation.

Since 2015, NVIDIA GPUs support fine-grained measurement of GPU
performance using Program Counter (PC) sampling~\cite{cuptipcsampling}.
Intel's GT-Pin~\cite{kambadur2015fast} and NVIDIA's NVBit~\cite{villa2019nvbit} provide APIs to instrument GPU machine code to collect fine-grained metrics.
While tools such as 
MAP~\cite{january2015allinea}, 
nvprof~\cite{nvprof}, 
Nsight-compute~\cite{nsightcompute},
TAU~\cite{shende2006tau}, 
and VTune~\cite{reinders2005vtune} use PC sampling or instrumentation
to associate fine-grained metrics with individual source lines for GPU
code, they do not associate metrics with loop nests or calling contexts for GPU device functions, which are important to understand the performance of complex GPU kernels.
For example, a template-based dot product kernel in the RAJA performance suite~\cite{rajaperf} yields 25 different GPU functions that implement the computation. 

To address these challenges, we are extending Rice University's HPCToolkit performance tools~\cite{adhianto2010hpctoolkit} to support scalable measurement and analysis of GPU-accelerated applications running on NVIDIA, AMD, and Intel GPUs.
HPCToolkit collects call path profiles and presents them with a graphical user interface that provides both profile  and trace views.
After our initial extensions to support GPU-accelerated programs, 
HPCToolkit has the following capabilities:

\begin{itemize}
\item It uses a GPU-independent measurement framework to  monitor and attribute performance of GPU code.

\item It employs wait-free queues for efficient coordination between application, runtime, and tool threads.

\item It supports measurement and attribution of fine-grained metrics using PC sampling and instrumentation. 
\item It employs compact sparse representations of performance metrics to support efficient collection, storage, and inspection of performance metrics within and across processes, threads, and GPU streams.
\item It employs a combination of distributed-memory parallelism and multithreading to aggregate global performance metrics across a large number of profiles.
\item It provides useful information to guide performance
  optimization, including heterogeneous calling contexts, derived metrics, and idleness analysis.
\end{itemize}

We present some early experiences with codes
from the Exascale Computing Project to illustrate HPCToolkit's
attribution of fine-grained measurements to heterogeneous calling contexts
on multiple GPU platforms and its capability to measure and analyze executions across hundreds of GPUs.

This rest of the paper is organized as follows.
Section~\ref{sec:related work} reviews related work and highlights HPCToolkit's features.
Section~\ref{sec:overview} describes HPCToolkit's workflow for GPU-accelerated applications.
Section~\ref{sec:measurement} describes the design of HPCToolkit's measurement framework for collecting GPU performance metrics.
Section~\ref{sec:program structure} discusses analysis of GPU binaries
for performance attribution.
Section~\ref{sec:analysis} presents scalable algorithms for
aggregating performance data from parallel programs.
Section~\ref{sec:user interfaces} describes HPCToolkit's profile and
trace views for analyzing measurements of GPU-accelerated
applications.
Section~\ref{sec:case studies} illustrates HPCToolkit's capabilities
with views of several codes from the Exascale Computing Project.
Section~\ref{sec:conclusion} reflects on our experiences and briefly outlines some future plans.

\section{Related Work}~\label{sec:related work}
Developing performance tools for GPU-accelerated applications has been the focus of considerable past and ongoing work.
NVIDIA provides tools~\cite{nsightsystem, nsightcompute, nvprof} to present a trace view of GPU kernel invocations and a profile view for individual kernels.
Intel's VTune~\cite{reinders2005vtune} monitors executions on both CPUs and GPUs.
AMD provides ROCProfiler~\cite{rocprofiler} to monitor GPU-accelerated applications.
In addition, a collection of third-party performance tools have been developed  for GPU-accelerated applications.
Malony et al.~\cite{malony2011parallel} describe early tools for collecting kernel timings and hardware counter measurements for CUDA and OpenCL kernels.
Welton and Miller~\cite{welton2019diogenes} investigated hidden performance issues that impact several HPC applications but are not reported by tool APIs.
Kousha et al.~\cite{kousha2019designing} developed a tool for monitoring communications on multiple GPUs.
Unlike the aforementioned tools, HPCToolkit collects call path profiles and shows calling context information in both trace and profile views.
Early work on HPCToolkit~\cite{chabbi2013effective} 
describes using GPU events and hardware counters for kernel-level
monitoring on NVIDIA GPUs to compute profiles that blame CPU code for associated GPU idleness.

With the increased complexity of GPU kernels, fine-grained measurement of performance metrics within GPU kernels are critical for providing optimization insights.
At present, only NVIDIA GPUs support using PC sampling~\cite{cuptipcsampling} to collect fine-grained instruction stall information. 
NVIDIA's nsight-compute collects data using PC sampling and provides performance information at the GPU kernel level.
CUDABlamer~\cite{cudablamer-protools19} was a proof-of-concept prototype that collects PC samples and reconstructs static call paths on GPUs with information from LLVM-IR.
Unlike CUDABlamer, HPCToolkit reconstructs GPU calling context trees
by analyzing GPU binaries and distributes costs of GPU functions based on PC sample counts.

Several vendor tools support instrumentation of GPU kernels.
NVIDIA's NVBit~\cite{villa2019nvbit} and Sanitizer
API~\cite{sanitizer}, as well as Intel's
GT-Pin~\cite{kambadur2015fast} provide callback APIs to inject
instrumentation into GPU machine code.
Tools can use these APIs to collect fine-grained metrics.
For example, Goroshov et al.~\cite{gorshkov2019gpu} use
instrumentation to measure basic block latency and detect hot code
regions.
GVProf~\cite{gvprof} instruments GPU memory instructions to profile value redundancies. 
In HPCToolkit, we use GT-Pin to measure instruction counts within GPU
kernels.

\sloppy
Scalable analysis of performance measurements will be critical for
gaining insight into executions on forthcoming exascale platforms.
NVIDIA's NVProf~\cite{nvprof} and Intel's
VTune~\cite{reinders2005vtune} record measurements as
traces. To our knowledge, these tools lack
support for scalable analysis of measurement data.
Scalasca, TAU~\cite{shende2006tau} and Vampir~\cite{knupfer2008vampir}
present data gathered by the Score-P measurement
infrastructure~\cite{ScoreP}. At execution finalization, Score-P
aggregates profile data in parallel into the CUBE storage format.
To date, there has only been a preliminary study exploring the addition of
sparsity to CUBE~\cite{CUBE}; for GPU-accelerated applications, sparsity is essential.
MAP~\cite{january2015allinea} selects a user-defined subset of collected samples at runtime to limit the amount of measurement data collected per thread.
This is effective for scalable overview analysis, however this does not retain sufficient data for in-depth analyses.

Past research has used trace analysis to  identify performance bottlenecks within and across compute nodes.
Wei et al.~\cite{wei2020using} describe a framework that diagnoses scalability losses in programs using multiple MPI processes and CPU threads.
Choi et al.~\cite{choi2020end} analyze traces from simulators to estimate performance on GPU clusters.
Schmitt et al.~\cite{schmitt2017scalable} use Vampir~\cite{knupfer2008vampir}'s instrumentation of MPI primitives to
gather communication traces. From these traces, they construct a dependency graph and explore dependencies between communication events and GPU computations.
Unlike other tracing tools, HPCToolkit gathers CPU traces using sampling rather than instrumentation, which has much lower overhead.

\section{Overview}~\label{sec:overview}
\begin{figure}[t]
\centering
\includegraphics[width=0.47\textwidth]{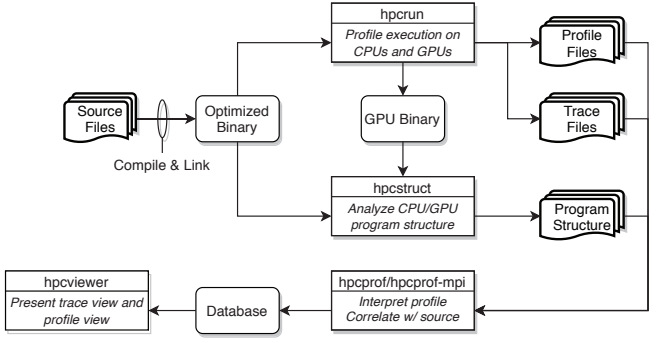}
\caption{HPCToolkit's workflow for analysis of GPU-accelerated applications.}
\label{fig:hpctoolkit}
\end{figure}

Figure~\ref{fig:hpctoolkit} shows HPCToolkit's workflow to analyze programs running on GPUs.
HPCToolkit's \textit{hpcrun} measurement tool collects GPU performance metrics using profiling APIs from GPU vendors or custom hooks with \texttt{LD\_PRELOAD}.
\textit{hpcrun} can measure programs that employ one or more GPU
programming models, including OpenMP, OpenACC, CUDA, HIP, OpenCL, and DPC++.
As GPU binaries are loaded into memory, \textit{hpcrun} records them for later analysis.
For GPUs that provide APIs for fine-grained measurement, \textit{hpcrun} can collect
instruction-level characterizations of GPU kernels using hardware
support for sampling or binary instrumentation.
\textit{hpcrun}'s output includes profiles and optionally traces.
Each profile contains a calling context tree in which each node is associated with a set of metrics.
Each trace file contains a sequence of events on a CPU thread or a GPU stream with their timestamps.

\textit{hpcstruct} analyzes CPU and GPU binaries to recover static
information about procedures, inlined functions, loop nests, and
source lines. There are two aspects to this analysis: (1) recovering
information about line mappings and inlining from compiler-recorded
information in binaries, and (2) analyzing machine code to recover
information about loops.

\sloppy
\textit{hpcprof} and \textit{hpcprof-mpi} correlate performance
metrics for GPU-accelerated code with program structure.
\textit{hpcprof} employs a multithreaded streaming aggregation
algorithm to quickly aggregate profiles, 
reconstruct a global calling context tree, and relate measurements
associated with machine instructions back to CPU and GPU source code.
To accelerate analysis of performance data from extreme-scale executions, \textit{hpcprof-mpi} additionally
employs distributed-memory parallelism for greater scalability.
Both \textit{hpcprof} and \textit{hpcprof-mpi} write sparse
representations of their analysis results in a database.

\begin{figure*}[t]
\centering
\includegraphics[width=0.75\textwidth]{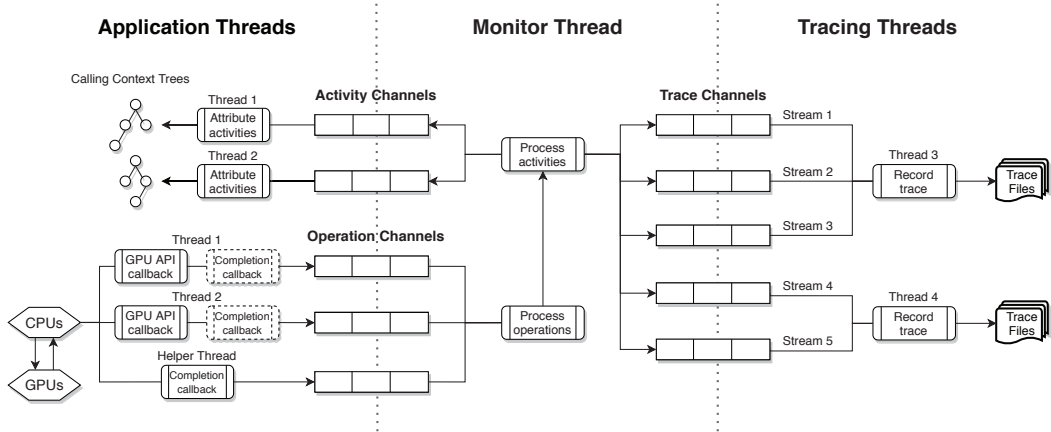}
\caption{HPCToolkit's infrastructure for coordinating application threads, monitor thread, and tracing threads.}
\label{fig:infrastructure}
\end{figure*}

Finally, \textit{hpcviewer} interprets and visualizes the database.
In its profile view, \textit{hpcviewer} presents a heterogenous
calling context tree that spans both CPU and GPU contexts, annotated
with measured or derived metrics to help users assess code performance
and identify bottlenecks.
In its trace view, \textit{hpcviewer} identifies each CPU or GPU trace
line with a tuple of metadata about the hardware (e.g.,
node, core, GPU) and software constructs (e.g., rank, thread, GPU stream)
associated with the trace.
Automated analysis of traces can attribute GPU idleness to CPU code.

\section{Performance Measurement on GPUs}~\label{sec:measurement}
HPCToolkit's \textit{hpcrun} collects GPU performance metrics and associates them with calling context at every GPU API invocation.
Section~\ref{subsec:infrastructure} describes HPCToolkit's unified infrastructure
for collecting and attributing performance metrics on AMD, Intel, and NVIDIA GPUs.
Section~\ref{subsec:fine-grained} describes how HPCToolkit collects
fine-grained metrics using hardware instruction sampling or binary instrumentation. 
Section~\ref{subsec:counters} describes support for measuring
GPU kernel executions with hardware counters.
Section~\ref{subsec:interaction} describes how HPCToolkit employs performance measurement substrates from GPU vendors.
Section~\ref{subsec:raw} describes how HPCToolkit collects metrics at
runtime and computes derived metrics during post-mortem analysis.  
Section~\ref{subsec:sparse} describes HPCToolkit's use of sparse
representations of performance metrics at runtime and as the products
of post-mortem analysis.
Section~\ref{subsec:combining} explains the utility of
combining measurements from multiple runs.

\subsection{Infrastructure}~\label{subsec:infrastructure}
Figure~\ref{fig:infrastructure} illustrates how {\em hpcrun} monitors
the execution of GPU-accelerated applications. As application threads
offload computations to GPUs, HPCToolkit employs a GPU monitor thread 
to asynchronously process measurement data from the GPUs. 
If tracing is enabled, \textit{hpcrun} creates one or more {\em tracing threads}
to record an activity trace for each GPU stream.

When an application thread performs an invocation $\mathcal{I}$ of a GPU operation (e.g., a
kernel or a data copy), {\em hpcrun} unwinds the application thread's 
call stack to determine the 
CPU calling context of $\mathcal{I}$, inserts
a placeholder $\mathcal{P}$ representing the operation in that context, 
communicates $\mathcal{I}$ and $\mathcal{P}$ it to the monitor thread, and initiates the GPU operation
after tagging it with $\mathcal{I}$.
The monitor thread collects measurements consisting of one or more GPU activities
$\mathcal{A}_1, \ldots, \mathcal{A}_n$ associated with $\mathcal{I}$
and
sends them back to the application thread for attribution below
$\mathcal{P}$ to form a heterogeneous calling context. 
When tracing is enabled, the monitor thread separates GPU activities
by their associated stream and sends each stream of activities to a
tracing thread. Each tracing thread records one or more GPU
streams of activities and their timestamps into trace files.
For efficient inter-thread communication, HPCToolkit uses
bidirectional channels, each consisting of a pair of wait-free
single-produer and single-consumer queues~\cite{zhouics20}.
The precise instantiation of HPCToolkit's monitoring infrastructure
is tailored to each GPU vendor's software for monitoring
GPU computations.

When using NVIDIA's CUPTI~\cite{cupti} and AMD's
ROCTracer~\cite{roctracer} libraries for monitoring GPU activities,
a monitor thread created by these libraries receives measurements of
GPU activities via a buffer completion callback. 
Each application thread $\mathcal{T}$ shares two channels with the GPU monitor
thread, including 
an activity channel $\mathcal{C_{A}}$, from which $\mathcal{T}$ receives
information about GPU activities associated with operations it invoked, and
an operation channel $\mathcal{C_{O}}$ on which $\mathcal{T}$ 
enqueues GPU operation tuples of $\mathcal{(I, P, C_{A})}$,
representing an invocation $\mathcal{I}$, its associated placeholder
$\mathcal{P}$, and its activity channel $\mathcal{C_{A}}$.
Every time the GPU monitor thread receives a buffer completion
callback, it drains its incident operation channels prior to
processing a buffer full of GPU activities.
The GPU monitor thread matches
each GPU activity $\mathcal{A}$, tagged with its invocation
$\mathcal{I}$, with its associated operation
tuple $\mathcal{(I, P, C_{A})}$. The monitor thread enqueues a pair
$\mathcal{(A, P)}$ into activity channel $\mathcal{C_{A}}$ to
attribute the GPU activity back to $\mathcal{T}$.

When using OpenCL~\cite{stone2010opencl} and Level
Zero~\cite{levelzero}, depending upon the GPU operation invoked,
either an application thread or a runtime 
thread will receive a completion callback providing measurement data.
At each GPU API invocation $\mathcal{I}$ by an application thread $\mathcal{T}$, {\em hpcrun} provides a
\texttt{user\_data} parameter~\cite{opencl-callback}, which includes 
a placeholder node $\mathcal{P}$ for the invocation
$\mathcal{I}$) and $\mathcal{T}$'s activity channel
$\mathcal{C_{A}}$. The OpenCL or Level Zero
runtime will pass \texttt{user\_data} to the completion callback
associated with $\mathcal{I}$.
At each completion callback, some thread receives measurement data
about a GPU activity $\mathcal{A}$.
Using information from its \texttt{user\_data} argument, the 
completion callback 
correlates $\mathcal{A}$ with
placeholder $\mathcal{P}$ and then
enqueues an operation of $\mathcal{(A, P, C_{A})}$ for
the monitor thread in its operation channel $\mathcal{C_{O}}$.
The monitor thread enqueues an $\mathcal{(A, P)}$ pair in $\mathcal{T}$'s activity channel $\mathcal{C_{A}}$.
If the thread receiving the callback enqueued $\mathcal{(A, P)}$ pairs
directly into $\mathcal{T}$'s activity channel $\mathcal{C_{A}}$,
$\mathcal{C_{A}}$ would need to be a multi-producer queue since
more than one thread may receive completion callbacks for $\mathcal{T}$.
Our design, which employs a GPU monitor thread created by {\em
  hpcrun}, replaces the need for a multi-producer queue with 
several wait-free single producer queues.

When tracing is enabled, the monitor thread checks the GPU stream id
$\mathcal{S}$ of each GPU activity and enqueues the activity and its
placeholder $\mathcal{P}$ into a trace channel for
$\mathcal{S}$. One or more tracing threads handle the recording of
traces.
Each tracing thread handles a set of trace channels by polling each
channel periodically and processing its activities.
For each activity in a trace channel for stream $\mathcal{S}$, the tracing thread records its
timestamp and placeholder in a trace file for $\mathcal{S}$.
Depending on the number of application threads used, the number of
tracing threads can be adjusted by users to balance tracing
efficiency with tool resource utilization.

\subsection{Fine-grained Performance Measurement}~\label{subsec:fine-grained}
On NVIDIA GPUs, HPCToolkit uses PC sampling to
monitor both instruction execution and stalls in GPU kernels.
On Intel GPUs, HPCToolkit uses Intel's GT-Pin to instrument GPU
kernels to collect fine-grain, instruction-level measurements.
AMD GPUs currently do not support either instrumentation-based or
hardware-based fine-grained measurement.

If PC sampling is used, the monitor thread receives a buffer full of
PC samples in a completion callback.  Each PC sample for a kernel includes an
instruction address, a stall reason, and a count of the times the
instruction was
observed.  The monitor thread enqueues an instruction measurement record into the
activity channel of the application thread that launched the
kernel. When an application thread receives an instruction measurement
record, it creates a node in its calling context tree representing
the GPU instruction as a child of the placeholder node for the
corresponding kernel invocation.

If instrumentation is used, when a GPU binary is loaded, HPCToolkit
injects code into each GPU kernel to collect measurements.
Measurement data is collected on a GPU and provided to HPCToolkit in a completion callback.
On Intel GPUs, HPCToolkit instruments a GPU kernel to count the execution frequency of each basic block.
In a completion callback following kernel execution, HPCToolkit
iterates over each basic block and propagates its execution count
to each instruction in the block.
Information about each instruction
and its count is sent to the monitor thread in an operation
channel. The monitor thread passes the
information back to the application thread that launched the kernel
using an activity channel. 
The application thread processes the instruction measurement like a PC sample.

\subsection{Measuring Performance with Hardware Counters}~\label{subsec:counters}
HPCToolkit uses hardware performance counters to observe how an
application interacts with an accelerated compute node. CPUs and GPUs each provide a collection
of programmable hardware counters that can be configured to measure device metrics (e.g., temperature and power), 
functional unit utilization, memory hierarchy activity, inefficiency, and more.
 
On CPUs, HPCToolkit uses the Linux {\tt perf\_event}
interface~\cite{weaver2013linux} to configure hardware counters with events and thresholds. 
HPCToolkit unwinds the call stack to attribute a metric to a call path each time a counter reaches a specified threshold. 
With appropriately chosen event thresholds, such measurement has low overhead.

On GPUs, HPCToolkit uses the University of Tennessee's
PAPI~\cite{terpstra2010collecting} as a vendor-independent interface to 
measure GPU activity using hardware counters. 
PAPI supports hardware counter-based measurement on NVIDIA, AMD, and Intel GPUs.
At present, the only way tools can associate hardware counter measurements with individual GPU kernels
using existing vendor APIs
is to serialize kernels and read data from counters before and after kernel
execution. Serializing kernels may both slow execution and alter execution behavior.
	
\subsection{Interaction with Measurement Substrates}~\label{subsec:interaction}
While developing HPCToolkit's GPU measurement infrastructure,
we encountered a few problems using each vendor's measurement
substrate(s). This section describes some the difficulties encountered 
and how they were handled.

Each GPU vendor and/or runtime system provides different levels of monitoring support.
NVIDIA's CUPTI~\cite{cupti} supports both coarse-grained and fine-grained measurements for CUDA programs.
AMD's ROCTracer~\cite{roctracer} only supports coarse-grained measurements for HIP programs.
Both of these monitoring frameworks enable a tool to register a
callback function that will be invoked at every GPU API
invocation. These callbacks can be used gather information about an
invocation, such as its calling context.
Intel's GT-Pin enables tools to add
instrumentation for fine-grained measurement of GPU kernels; however,
neither its OpenCL~\cite{stone2010opencl} or Level
Zero~\cite{levelzero} runtimes provide APIs for collecting
coarse-grained metrics. As a result, HPCToolkit wraps Intel's OpenCL and Level Zero
APIs using \texttt{LD\_PRELOAD} to collect custom
information in each API wrapper. Wrapping APIs is sensitive to changes
in APIs as the runtimes evolve (interfaces in Level Zero have changed over
the last few months) and may not provide access to all information of
interest, e.g., implicit data movement associated with kernel arguments
in OpenCL.

As a GPU program executes, vendor runtime and/or tool APIs
typically create helper threads.
For example, if PC sampling is used, CUPTI creates a short-lived
helper thread each time the application launches a kernel.
Thus, in a large-scale execution that launches kernels millions of times,
CUPTI will create millions of short-lived threads.
Similarly, several components in AMD's software stack create threads,
including the HIP runtime, ROCm debug library, and ROCTracer.
To reduce monitoring overhead, HPCToolkit wraps \texttt{pthread\_create} to
check if a thread is created by a GPU runtime or its tool API.
If yes, HPCToolkit avoids monitoring the thread.

In CUPTI and ROCTracer, a single helper thread in each process handles
GPU measurement data using the buffer completion callback.
However, for OpenCL and Level Zero, an event completion callback may
be asynchronous, as described for OpenCL~\cite{opencl-callback}.
Hence, it is \textit{hpcrun}'s responsibility to ensure that
callbacks gather and report information in a thread-safe fashion.
To avoid races reporting data back to an application thread,
\textit{hpcrun} first delivers measurement data from the thread
that receives the callback, which might be the application thread, to a
monitoring thread using a point-to-point operation channel between the
threads. The monitoring thread then delivers measurement data back to the proper
application thread using a point-to-point activity channel. 

While CUPTI and ROCTracer typically order activities within each stream, 
the order in which GPU activities are reported is undefined for OpenCL~\cite{opencl-callback}.
On Power9 CPUs, we have even observed observed overlapping intervals on a
stream using CUPTI. 
Rather than taking extreme measures to order each stream's
activities in \textit{hpcrun}, we simply record each stream 
into a trace file and note if any activity is added out of
order. If so, HPCToolkit sorts the trace stream to correct the order 
during post-mortem analysis.

\subsection{Measuring and Computing Metrics}~\label{subsec:raw}
As a GPU-accelerated program executes, HPCToolkit collects performance metrics and
associates them with heterogeneous calling contexts. 
HPCToolkit supports several strategies for measuring and computing metrics.
A {\em raw} CPU or GPU metric for a heterogeneous calling
context in an application thread is simply the sum of all measured
values of a specific kind associated with that context. 
For instance, raw metrics for GPU data copies associated with
a context include the operation count, total bytes copied, and total copy time. 

To facilitate analysis, HPCToolkit also computes two types of {\em
  derived} metrics.  The first type of derived metrics is computed
during post-mortem analysis by HPCToolkit's {\em hpcprof}. Built-in
derived metrics for combining metrics from different thread profiles
during post-mortem analysis include {\em sum}, {\em min}, {\em mean},
{\em max}, {\em std.\ deviation}, and {\em coefficient of variation}.
With the exception of {\em sum}, these metrics can provide insight into imbalances across threads.
The second type of derived metrics is computed in HPCToolkit's {\em
  hpcviewer} user interface. HPCToolkit uses {\em hpcviewer} to
compute GPU metrics including GPU utilization and 
GPU theoretical occupancy.

Computing some GPU metrics requires a bit of creativity.
For instance, NVIDIA's CUPTI reports static information about a
kernel's resource consumption (e.g., registers used) each
time a kernel is invoked. To avoid the need for a special mechanism
for collecting such metrics, HCToolkit simply records odd raw metrics such as the 
the sum of the count of registers used over all kernel invocations 
in a particular calling context and the count of kernel
invocations in that context. After summing these raw metrics over
threads and MPI ranks, {\em hpcviewer} computes the ratio of these two
values to recover the number of registers used. 

\subsection{Sparse Representation of Metrics}~\label{subsec:sparse}
\textit{hpcrun} maintains a Calling Context Tree (CCT) for each CPU
thread or GPU stream it measures.
In a CCT, each node represents the address of a machine instruction in a CPU or GPU binary as a (load module, offset) pair.
When a CCT node is allocated, it receives a companion metrics array to
store associated performance metrics. In HPCToolkit, well over 100
metrics can be measured; some for CPUs and some for GPUs.
When measuring the performance of GPU-accelerated programs,
many CCT nodes have CPU metrics only; all of their many GPU metrics are zero.
Storing zero values for all unused metrics at a CCT node would waste
considerable memory.

To reduce storage during measurement, {\em hpcrun} partitions metrics into {\em kinds}, such as {\em GPU kernel
  info} kind, {\em GPU instruction stall} kind, and {\em CPU time} kind. 
Each CCT node is associated with a metric kind list, and each metric
kind represents an array of one or more metrics.
For example, when measuring an execution with PC sampling, the CCT
node representing a GPU kernel has {\em GPU kernel info} kind and {\em GPU
  instruction sampling info} kind.
The kernel kind includes kernel running time, register usage, and shared
memory usage, among others.

\begin{figure}[t]
\centering
\begin{subfigure}[h]{0.4\textwidth}
         \centering
         \includegraphics[width=\textwidth]{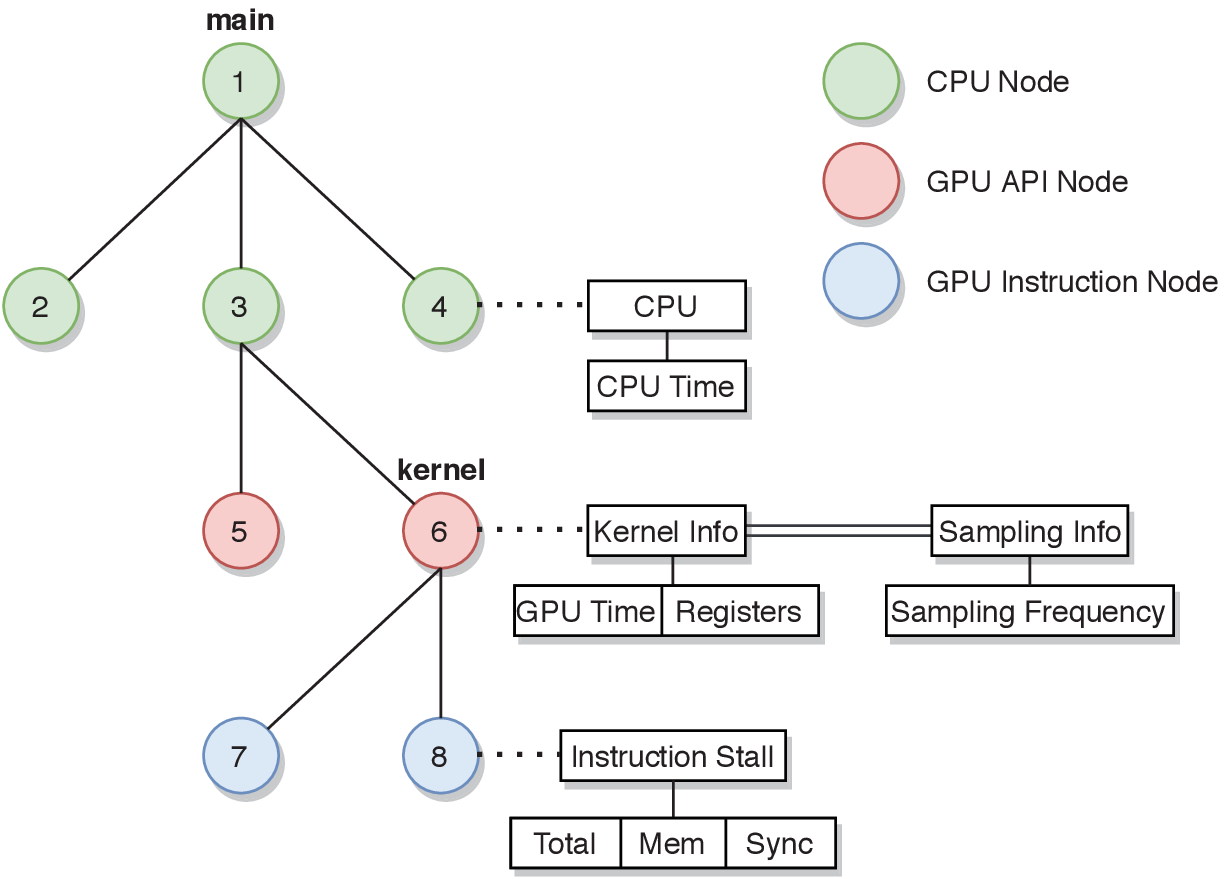}
         \caption{Sparse internal representation}
         \label{fig:sparse-internal}
         ~\\
\end{subfigure}
\begin{subfigure}[h]{0.3\textwidth}
         \centering
         \includegraphics[width=\textwidth]{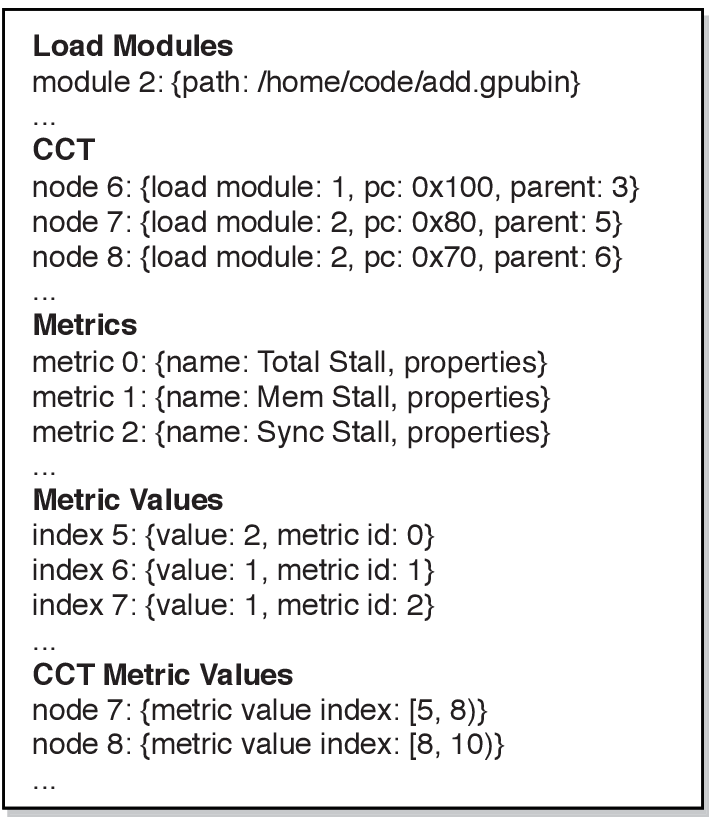}
         \caption{Sparse output format}
         \label{fig:sparse-output}
\end{subfigure}
\caption{hpcrun's sparse representation of a CCT and its metrics in memory and on the disk}
\label{fig:sparse-hpcrun}
\end{figure}

Figure~\ref{fig:sparse-internal} illustrates the sparse
representation of metrics associated with CCT nodes as {\em hpcrun}
measures the performance of a GPU-accelerated application.
In the figure, each CCT node is categorized as a CPU node, a GPU API
node, or GPU instruction node. Each type of CCT node is associated with different metric kinds.

In addition to representing metrics sparsely in memory, {\em hpcrun} also writes
profiles to the file system using a sparse format to save space.
The output format of each profile file has the following sections.
\begin{itemize}
\item A \textit{Load Modules} section that contains all the libraries loaded in execution.
\item A \textit{CCT} section that depicts the structure of a CCT, including each node's module id, offset, and parent fields.
\item A \textit{Metrics} section that contains the index and name of
  each performance metric, as well as some properties.
\item A \textit{Metric Values} section and a \textit{CCT Metric
  Values} section that indicates the metric values associated with
  each CCT node.
\end{itemize}

To generate the \textit{Metric Values}, {\em hpcrun} 
iterates through the metrics kind list of each CCT node, counts the
number of non-zero metrics $N$, and records their values.
In the \textit{CCT Metric Values} section, a CCT node with an index range $[I, N)$ indicates that it
has metrics in the \textit{Metric Value} section 
at positions from $I$ to $I + N - 1$.
Profiles produced by {\em hpcrun} employ this scheme to represent only
non-zero metrics. 

Figure~\ref{fig:sparse-output} illustrates the sparse representation
of metrics in {\em hpcrun}'s output files.
In the \textit{CCT Metric Values} section of the figure, node 7 has three metrics---metric index 5, metric index 6, and metric index 7.
We locate metric index 5's value (2) by at index (5) in the \textit{CCT Metric Values} section.
Further, we can retrive the metrics's name (Total Stall) by looking up
its \textit{metric id} (0) in the \textit{Metrics} section.

\subsection{Minimizing Measurement Errors}
\label{subsec:combining}

There are many ways that HPCToolkit can measure the performance of a
GPU-accelerated application: sampling on the CPU using timers or
hardware counters, reading GPU hardware counters before and after GPU
operations, profiling and/or tracing timings of GPU operations, and
using PC sampling or instrumentation to collect fine-grained
measurements of GPU kernels. Performing all of these measurements in a
single execution is unwise. For instance, fine-grained instrumentation
of GPU kernels can distort GPU profiles and traces recorded in the
same execution. To minimize the distortion in measurements, it is best
to collect each kind of measurements in a separate run, if one can
afford multiple runs. HPCToolkit's post-mortem analysis can combine
performance measurements from multiple runs to produce a comprehensive
representation of an application's performance.

\section{Program Structure Recovery}~\label{sec:program structure}
Relating performance metrics associated with a node in a calling
context tree to source code requires understanding the mapping from
machine instructions to source lines, loops, functions, and inlined
code.  HPCToolkit's {\em hpcstruct} analyzes both CPU and GPU binaries to determine this
mapping. This strategy works for any statically-compiled language,
including C, C++, and Fortran.


Binaries for GPU-accelerated applications are rather unusual. A host
application typically has GPU binaries embedded within. Furthermore,
NVIDIA and Intel GPU binaries are unlike any CPU binaries we have ever
seen. While each of the GPU vendors rely in part on ELF
representations for GPU code, their formats for GPU binaries are quite
different. 

NVIDIA's GPU binaries sometimes have device functions embedded inside
a global GPU function. Each function is in a separate text segment and all functions start with offset 0 in its symbol table.
To aid measurement and analysis, we relocate each function to a unique address—its offset in the binary’s section table.
We split any overlapping GPU functions into disjoint address ranges.
Then we use NVIDIA’s nvdisasm binary tool to analyze GPU machine code and dump a control flow graph for each function in a binary.
Because nvdisasm renders merged basic blocks in control flow graphs, we split superblocks into basic blocks to fit analysis.

Intel's GPU binaries are embedded in a ``fat'' binary in which each
section contains information about a single GPU kernel.
At runtime, we dump the fat binary to the file system. During post-mortem
analysis, {\em hpcstruct} reads a
fat binary, iterate through all kernels, and processes the ELF binary
for each kernel separately. It use Intel's IGA API to decode each kernel
binary to obtain its control flow graph.

AMD's GPU binaries are also embedded in a ``fat'' binary. The locations of AMD's GPU binaries inside the fat binary are specified by
Uniform Resource Identifiers (URIs) reported by ROCm debug API and
ROCTracer. AMD's GPU binary uses a conventional ELF representation.
We use the Dyninst~\cite{dyninst} binary analysis framework to decode AMD GPU instructions and construct control flow graphs.
The Dyninst team has developed preliminary support for analyzing AMD
GPU binaries, including support for decoding scalar instructions and
resolving direct control flow; support for decoding vector
instructions and resolving indirect control flow is a work in progress.
 
Control flow graphs extracted from Intel and NVIDIA GPU binaries
are converted to a uniform representation and injected into Dyninst,
which analyzes loop nests they contain.
We classify GPU instructions by their operation types, instruction length, and kind of memory they access.
As Dyninst can analyze AMD GPU binaries, the aforementioned analyses
are performed directly in Dyninst.

{\em hpcstruct} uses Dynisnt to read DWARF sections in a GPU binary to attribute instructions to source lines and recover inlined functions.
For AMD, Intel and NVIDIA GPU binaries, it is worth noting their
information about inlined code is not perfect. None of the GPU
binaries provides information about the call site of an inlined
function, even when they are compiled with debug information (e.g.,\texttt{-lineinfo}) and optimization (e.g., \texttt{-O3}).

If \textit{hpcrun} records many GPU binaries at for a complex exeuction, \textit{hpcstruct} analyzes GPU binaries in parallel.

\section{Scalable Performance Analysis}~\label{sec:analysis}
This section describes our approach to improve the scalability of our performance analysis, to handle the additional data gathered from GPU-accelerated applications running at the extreme scales supported by forthcoming systems.
\cref{sec:streamingagg} describes our new approach for scalable performance
analysis of measurement data from extreme-scale executions.
\cref{sec:fmtdesign} presents the structure of the new sparse
data format HPCToolkit uses to address the sparsity in measurements
and analysis results for heterogenous applications.
\cref{sec:gpucct} describes an algorithm to reconstruct approximate
GPU calling context trees for complex GPU kernels. 

\subsection{Streaming Aggregation}\label{sec:streamingagg}
Forthcoming exascale compute platforms pose significant challenges for performance tools.
Performance issues in applications or system software may only become
apparent during executions at very large scales, requiring performance
tools to support measurement of applications executing on systems with
tens of thousands of compute nodes equipped with multicore processors
and GPUs.
The aggregation of measurement data into statistics suitable for human inspection becomes increasingly expensive as the scale of executions increases, requiring tools to employ well-designed algorithms for data aggregation with good scaling to large inputs.

To address these issues, we developed a novel algorithm for
performance analysis that exploits both process-based (via MPI ranks)
and thread-based parallelism. Exploiting thread-level parallelism
reduces communication overheads and the memory footprint during analysis. 
The key to our algorithm is to process profiles of individual application threads or GPU streams as concurrent parallel tasks, sharing a limited set of data structures between profiles containing the resulting aggregated values.
In effect, we ``stream'' the input profile data in parallel to the appropriate destination, retaining only values required for the final ``aggregation'' of all inputs, hence the name.
Our algorithm then can be described as a sequence of operations applied in parallel on the profile inputs, as follows:

{\em Input Acquisition:}
Before any other processing is performed, the set of input profiles is
acquired and initial offsets are prepared to facilitate later data reads.
A profile contains a tree of call paths and measurements
attributed to nodes in call paths as described in
\cref{subsec:sparse}, and may additionally include an execution trace
referencing individual call paths.
These profiles are then distributed evenly across the available ranks, where they continue to be processed in parallel using a dynamic scheduling algorithm.

{\em Call Path Profile Unification:}
Once the profiles have been distributed across the available ranks, we
unify the tree of call paths from each profile into a single tree available at the root of a collective reduction operation.
This can be done optimally utilizing a reduction tree of the same arity as the number of threads available in each rank $t$, as the rank has sufficient capacity to handle data from $t$ children at once.

{\em Calling Context Expansion:}
As call path nodes are received in parallel at the root rank, both
from allocated profiles and child ranks, calling context nodes are
created based on program structure information for each instruction in a call path.
Basic program structure information (lines and inlined code) can be
generated from line map and inlining information recorded in a binary
by a compiler. If information about loops is of interest,   
then program structure files generated by {\em hpcstruct}, as descried
in Section~\ref{sec:program structure}, can be imported.
Once the unified calling context tree is fully constructed on the root
rank, the complete conversion mapping from call paths nodes to calling
contexts is broadcast to all other ranks, which use it to construct compatible subsets of the calling context tree represented by their contributions.
Compared to our previous implementation of post-mortem analysis using MPI-everywhere parallelism, 
our new implementation has a smaller memory footprint 
when using the same number of cores because multiple threads in each MPI rank share a single copy of the unified calling context tree.

{\em Statistic Generation:}
At this stage, the measurements present within each profile are read in parallel, propagating values up the calling context tree to generate per-thread metrics for every referenced calling context.
On each rank these metrics are then composed into a set of accumulators present on each calling context, which are used to generate metric statistics global to all threads in the application's execution.
At this point the per-thread metrics can be written to the output database immediately, while the accumulators are aggregated by a second reduction operation similar to the first.

{\em Trace and Final Outputs:}
If a profile has a companion execution trace, at this stage its trace
sequence is converted one sample at a time to reference calling
contexts instead of the call paths present at runtime and then written directly into the output database.
After all profiles have been processed, the final unified calling context tree and global statistics are written to the database by the root rank.

\subsection{Sparse Representations of Analysis Results}\label{sec:fmtdesign}
Inspired by \textit{Compressed Sparse Row} (CSR) format~\cite{CSR} for
storing a sparse matrix, we designed and implemented a pair of compact
sparse formats to store HPCToolkit's performance metrics for threads
and GPU streams. These sparse formats can save storage and ensure efficient access of values needed for presentation and inspection.

\begin{figure}[t]
  \centering
  \begin{tikzpicture}[scale=.8]
    \small
    \draw[thick] (0,0) -- +(.1,-.1)
          +(.2,-.2) coordinate (cm 4)
          +(0,0) -- coordinate[pos=.4] (cm 3 split) coordinate[pos=.68] (cm 2 split)
            node[auto,sloped] {Profile ID}
          ++(40:2.5) -- +(.1,-.1)
          +(.2,-.2) coordinate (cm 1);
    \path (cm 2 split) +(.25,-.25) node[rotate=40] {\large$\cdots$} (cm 3 split) +(.2,-.2) coordinate (cm 3);
    \foreach \coord in {cm 1,cm 3,cm 4} {
      \draw[fill=black!0!white,thick] (\coord) rectangle +(2.6,-2.6);
      \foreach \lenx in {0,.2,...,2.4} {
        \foreach \leny in {0,.2,...,2.4} {
          \pgfmathparse{(rnd > 0.8) ? 0.2 : 0} \edef\x{\pgfmathresult}
          \path[fill=black!30!white] (\coord) ++(\lenx,-\leny) rectangle +(\x,-\x);
        }
      }
      \foreach \len in {.2,.4,...,2.4} {
        \draw[color=black!40!white]
          (\coord) ++(\len,0) -- +(0,-2.6)
          (\coord) ++(0,-\len) -- +(2.6,0);
      }
      \draw[draw=black,thick] (\coord) rectangle coordinate[midway] (front center) +(2.6,-2.6);
    }
    \path (cm 1) -- node[auto] {Metric ID} +(2.6,0);
    \path (cm 4) ++(0,-2.6) -- node[auto,sloped] {Context ID} +(0,2.6);
    \node[fill=black!0!white, align=right] at (front center) {
      vals: $\lbrack\hspace{2em}\rbrack$ \\
      mids: $\lbrack\hspace{2em}\rbrack$ \\
      cidxs: $\lbrack\hspace{2em}\rbrack$
    };

    \draw[thick] (5.5,0) -- +(.1,-.1)
          +(.2,-.2) coordinate (cm 4)
          +(0,0) -- coordinate[pos=.4] (cm 3 split) coordinate[pos=.68] (cm 2 split)
            node[auto,sloped] {Context ID}
          ++(40:2.5) -- +(.1,-.1)
          +(.2,-.2) coordinate (cm 1);
    \path (cm 2 split) +(.25,-.25) node[rotate=40] {\large$\cdots$} (cm 3 split) +(.2,-.2) coordinate (cm 3);
    \foreach \coord in {cm 1,cm 3,cm 4} {
      \draw[fill=black!0!white,thick] (\coord) rectangle +(2.6,-2.6);
      \foreach \lenx in {0,.2,...,2.4} {
        \foreach \leny in {0,.2,...,2.4} {
          \pgfmathparse{(rnd > 0.8) ? 0.2 : 0} \edef\x{\pgfmathresult}
          \path[fill=black!30!white] (\coord) ++(\lenx,-\leny) rectangle +(\x,-\x);
        }
      }
      \foreach \len in {.2,.4,...,2.4} {
        \draw[color=black!40!white]
          (\coord) ++(\len,0) -- +(0,-2.6)
          (\coord) ++(0,-\len) -- +(2.6,0);
      }
      \draw[draw=black,thick] (\coord) rectangle coordinate[midway] (front center) +(2.6,-2.6);
    }
    \path (cm 1) -- node[auto] {Profile ID} +(2.6,0);
    \path (cm 4) ++(0,-2.6) -- node[auto,sloped] {Metric ID} +(0,2.6);
    \node[fill=black!0!white, align=right] at (front center) {
      vals: $\lbrack\hspace{1.5em}\rbrack$ \\
      pids: $\lbrack\hspace{1.5em}\rbrack$ \\
      midxs: $\lbrack\hspace{1.5em}\rbrack$
    };
  \end{tikzpicture}  
  \caption{Visualization of Profile Major Sparse format (left) and CCT Major Sparse format (right). Each plane uses a modified CSR. Gray cells represent non-zeros on each plane.}
  \label{fig:sf}
\end{figure}
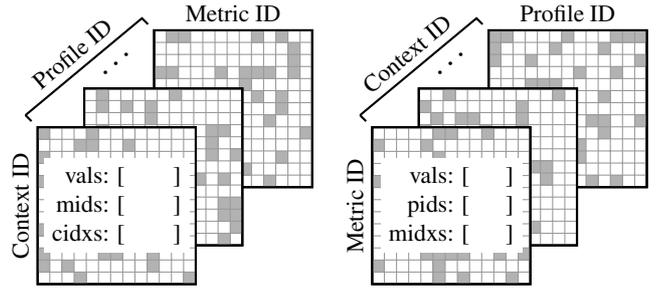

\cref{fig:sf} shows a high-level visualization of our design: \textit{Profile Major Sparse} (PMS) format and \textit{CCT Major Sparse} (CMS) format. As mentioned earlier, one of our goals is to access the data efficiently, which means accessing the data both within one profile and between profiles. Therefore, we designed PMS and CMS to compare performance within and across threads or GPU streams, respectively, so we only need to open one file for all comparisons as long as they belong to the same type.

Unlike CSR, we need to locate a value with three indices: a metric ID,
a context ID, and a profile ID. If we consider the matrix represented
by CSR a sparse plane, then our formats represent sparse cubes. The
filled cells represent non-zero values and the three arrays show how
to represent each plane. PMS consists of a vector of profile offsets,
one per profile, that indicate the start of each plane of (profile,
context, metric) triples. CMS consists of a vector of context offsets,
one per calling context, that indicate the start of each plane of
(context, metric, profile) triples. For brevity, in the rest of
\cref{sec:fmtdesign}, 
we only discuss the design details of CMS.

As \cref{fig:sf}, each plane can be seen as a CSR. $row$ in CSR is a
dense array, however, after some experiments, we found many repeating
values in $midxs$ for CMS, which means there are no non-zero values
for these metrics for the specific context, and we call them
\textit{empty} metrics. 
As discussed previously, we expect many empty metrics in calling
context trees for heterogeneous programs. 
To exploit this sparsity, we make $midxs$ a sparse array: each entry is a pair of metric ID and this metric's starting index ($midx$) in $pids$ and $vals$. 

With $x$ non-zeros, $m$ non-empty metrics in an average profile,
and $c$ contexts, then CMS uses \(\mathcal{O}(c \times (2x + m + 1))\)
space. This saves space when \((2x + m + 1) < MP\), where $M$ is the
number of metrics and $P$ is the number of profiles. \((2x + m + 1)\)
can even be slightly larger than $MP$ since the dense version uses a
consistent number of bytes for each data, but CMS can use fewer for
some data whenever appropriate, for instance, metric ID. To access a
specific value, it takes constant time to locate a context plane and
\(\mathcal{O}(\log m)\) time to binary search for the metric
index {\em midx}. To compare performance between profiles for a
specific context and metric, go to $pids[midx]$ and then compare
all of the values until the start of the next metric. 
To access one specific value, let $p$ be the number of profiles associated with the context and the metric, then it takes \(\mathcal{O}(\log m + \log p)\) time in total. Therefore, CMS can save storage and ensure efficient access of values. 

To construct our sparse formats using both distributed-memory and
multithreaded parallelism, {\em hpcprof-mpi} uses MPI for
communication between ranks and shared memory for communication
between OpenMP threads. Both CMS and PMS use the same general idea to
generate the file: find the right offset of each plane, collect the
related bytes, write the bytes to the offset, and record the location
in the offsets section. {\em hpcprof-mpi} uses \textit{exscan} operations within and
between ranks to find the right offsets, and then all threads can
mostly finish the remaining three steps concurrently without communication. 

For PMS, {\em hpcprof-mpi} splits work within and between ranks based on the
number of profiles to ensure load balance. However, since different
calling contexts may have huge differences in the number of associated
metrics, for CMS, {\em hpcprof-mpi} partitions
the work within and between ranks based on the number of non-zeros for
contexts to ensure load balance. As a performance tool for exascale
applications, {\em hpcprof-mpi} doesn't assume it has enough memory to
store all contexts or
profiles at the same time. For that reason, it processes the data in an
out-of-core fashion. For example, for CMS, {\em hpcprof-mpi} 
has a pre-set maximum memory that it can use for one round, and it processes the data in
multiple rounds if necessary. 

\subsection{GPU Calling Context Tree}\label{sec:gpucct}
To analyze the performance of complex GPU kernels that call device
functions, it is useful to organize performance data into GPU
calling context trees. 
Using binary instrumentation to collect instruction traces, we could
reconstruct a GPU calling context tree on the CPU as an application
executes.
However, this method would have high overhead with frequent communication between CPUs and GPUs to copy traces~\cite{gvprof}.

To address this, we designed a method to reconstruct approximate GPU
calling context trees offline from fine-grain instruction-level measurements
gathered using instrumentation or PC samples. This approach can be
applied in any environment where collecting precise call stacks is expensive.

Our method reconstructs an approximate GPU calling context tree for each GPU kernel invocation with the following four steps.
\begin{enumerate}
\item Construct a GPU static call graph based on function symbols and call instructions. Initialize weights on call edges using exact counts of call instructions or call instruction sample counts.
\item 
{\em For call graphs based on samples:} if a function has samples and none of its incoming call edges has a non-zero weight, we assign each of its incoming call edges a weight of one; we repeat this propagation through callers until at least one call edge of a function has a non-zero weight.
\item Identify strongly connected components (SCCs) using Tarjan's algorithm~\cite{tarjan1972depth}.
In the call graph, add an SCC node to represent the set of nodes in each SCC. 
Link external calls to functions inside an SCC to the SCC node and remove call edges between functions in the SCC.
\item Finally, build a calling context tree by splitting the call graph. Like Gprof~\cite{graham1982gprof}, assume that every invocation of a function takes the same time.
Apportion the number of samples or instructions in each function among its call sites using ratios of calls from each call site to the total number of calls from all call sites.
\end{enumerate}

\begin{figure}[t]
\centering
\includegraphics[width=0.35\textwidth]{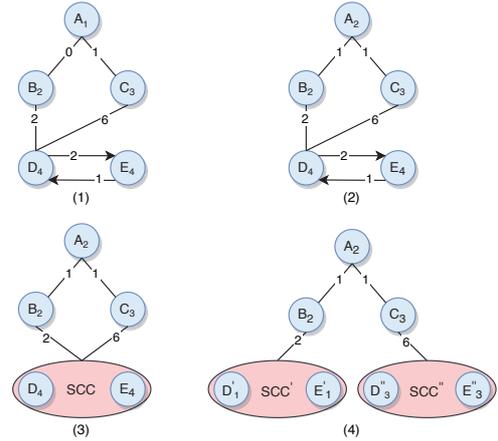}
\caption{An example of splitting a call graph into a calling context tree.
A-D denote functions.
Each subscript denotes the number of samples associated with the function.
Each edge is annotated with its number of call samples.}
\label{fig:gpu-cct}
\end{figure}

Figure~\ref{fig:gpu-cct} shows the reconstruction process for a small synthetic example.
We first construct a static call graph in Step 1.
In Step 2, we assign the edge between A to B one call sample because B does not have a sampled call site.
If executed instructions are collected, each edge representing a call that executed should have a non-zero weight.
In Step 3, we identify an SCC that contains D and E.
Finally, we apportion the number of samples of the SCC using ratios of
weights of calls from each call site to the total weight of calls from all call sites.


\section{User Interfaces}~\label{sec:user interfaces}
\textit{hpcviewer} is HPCToolkit’s graphical user interface (GUI),
built on top of the Eclipse Rich Client Platform (RCP), which enables
us to support multiple platforms including Windows, macOS, Linux x86/64
and Linux ppcle64. We plan to support Linux Arm in the near
future; at present, we are awaiting a forthcoming release of Eclipse
for Arm.
\textit{hpcviewer} is designed to support huge application performance
databases by using sparse data representations, loading GUI components
lazily, and employing multiple threads to read and load data
dynamically, as needed.  

The database generated by \textit{hpcprof} consists of 4 dimensions: \emph{profile}, \emph{time}, \emph{context}, and \emph{metric}.  
We employ the term \emph{profile} to include any logical threads (such as OpenMP, pthread and C++ threads), and also MPI processes and GPU streams.
The \emph{time} dimension represents the timeline of the program's execution, \emph{context} depicts the path in calling-context tree, and \emph{metric} constitutes program measurements performed by \textit{hpcrun} such as cycles, number of instructions, stall percentages and ratio of idleness.

To simplify performance data visualization, {\em hpcviewer} restricts
display two dimensions at a time: the \emph{profile viewer} displays
pairs of $\langle$context, metric$\rangle$ or $\langle$profile,
metric$\rangle$ dimensions; and the \emph{trace viewer} visualizes the
behavior of threads or streams over time.
Section~\ref{sec:profile-view} describes HPCToolkit's \emph{profile
  viewer} and Section~\ref{sec:trace-view} describes it \emph{trace viewer}. 

\subsection{Profile viewer}
\label{sec:profile-view}
HPCToolkit's profile viewer provides a code-centric user interface for interactive examination of performance databases. 
As shown in Figure~\ref{fig:quicksilver}, it displays pairs of $\langle$context, metric$\rangle$ dimensions, which enables users to view context-sensitive performance metrics correlated to program structure and mapped to a program’s source code, if available. 
It can also present an arbitrary collection of performance metrics gathered during one or more runs or compute derived metrics.

Measurements can be analyzed in a variety of ways: \emph{top-down} in a calling context tree, which associates costs with the full calling
context in which they are incurred; \emph{bottom-up} in a view that apportions costs associated
with a function to each of the contexts in which the function is called; and in a \emph{flat view} that
aggregates all costs associated with a function independent of calling context. 
This multiplicity of code-centric perspectives is essential to understanding a program’s performance for
tuning under various circumstances. 
\emph{hpcviewer} also supports a \emph{thread-centric} perspective, which enables one to see how a performance metric for a calling context differs across threads. 
The viewer can plot a graph of metric values associated with the selected node in CCT for all processes or threads.
This functionality allows users to display pairs of $\langle$profile, metric$\rangle$ dimensions.

To collect all necessary metrics, GPU performance tools often employ multiple runs and merge all metrics together. 
For example, Nsight-compute runs nine passes to collect all of its default metrics for a small GPU kernel; this approach is infeasible for a large-scale application.
As mentioned in Section~\ref{subsec:raw}, hpcviewer lets users compute derived metrics based on measurements gather using a single pass to identify optimization opportunities.
A derived metric is a spreadsheet-like formula composed from existing metrics, operators, functions, and numerical constants.
For example, for each GPU kernel, based on the number of total PC samples ($S$) and stalled PC samples ($S_{stall}$), we can estimate the Warp Issue Rate ($W$) of schedulers as $W = \frac{S - S_{stall}}{S}$. 


\subsection{Trace viewer}
\label{sec:trace-view}
HPCToolkit's trace viewer provides 
a time-centric user interface for the interactive examination of a sample-based time series (hereafter referred to as a trace) view of program execution. 
It is designed to interactively present traces of large-scale
execution across both CPUs and GPUs, relating activity to both
hardware contexts (e.g., nodes, GPUs, cores) and software contexts (e.g., MPI ranks, threads, GPU contexts, and streams).

As shown in Figure~\ref{fig:hpcviewer_callstack}, the trace viewer's main pane shows $\langle$profile, time$\rangle$ dimensions, for each available call-stack depth. 
By changing call-stack depth, a user can change the granularity of
trace lines, and gain insight into an execution at different levels of abstraction. 
For the routine pointed by the cursor, functions from the call stack are listed on the right to the main pane. 
Each routine is uniquely identified with a specific color, while idle activity is assigned the color white.

The trace viewer's \emph{Statistic} and \emph{GPU Idleness Blame} tabs
analyze the information in traces and offer some high-level characterizations of what the traces show.
The \emph{Statistic} tab calculates the percentage of the area
occupied by each routine in the main pane and lists routines in
descending order according to their percentage of the area. 
The \emph{GPU Idleness Blame} tab employs blame analysis in an attempt
to help application developers understand the causes of GPU idleness. 
To do so, it identifies times when all GPU streams are idle and at
least one CPU thread is active. In such cases, it partitions the cost
of GPU idleness among routines being executed by active CPU threads. 
The \emph{GPU Idleness Blame} tab then presents normalized blame associated with each CPU function in sorted in descending order.
CPU routines associated with high GPU idleness may be candidates for optimization. 


\begin{table}[t]
\caption{Experimental platforms}
\label{tab:platforms}
\footnotesize
\centering
\begin{tabular}{|c|c|c|}
\hline
\textbf{Node} & \textbf{Hardware}                                                                                        & \textbf{Software}                                                \\ \hline
OLCF Summit        & \begin{tabular}[c]{@{}c@{}}2$\times$IBM POWER9 CPUs\\ 6$\times$NVDIA V100 GPUs\end{tabular}              & \begin{tabular}[c]{@{}c@{}}GCC-6.4.0\\ CUDA-11.0.3\end{tabular}  \\ \hline
Argonne JLSE Iris   & \begin{tabular}[c]{@{}c@{}}1$\times$Intel E3-1585Lv5 CPU\\ 1$\times$Intel Iris Pro P580 GPU\end{tabular} & \begin{tabular}[c]{@{}c@{}}GCC-8.3.1\\ DPC++, oneAPI beta 10\end{tabular} \\ \hline
Local AMD     & \begin{tabular}[c]{@{}c@{}}2$\times$AMD EPYC 7402 CPUs\\ 2$\times$AMD MI50 GPUs\end{tabular}             & \begin{tabular}[c]{@{}c@{}}GCC-8.3.1\\ Rocm-3.8\end{tabular}     \\ \hline
\end{tabular}
\end{table}

\section{Case Studies}~\label{sec:case studies}
We tested HPCToolkit’s support for analyzing GPU-accelerated
applications on the Summit supercomputer with NVIDIA GPUs, the JLSE
Iris testbed which consists of nodes equipped with Intel Skylake
processors that have integrated GPUs, and a local machine with AMD GPUs.
The hardware and software specification of the three platforms are shown in Table~\ref{tab:platforms}.
We evaluated HPCToolkit with three HPC applications described below:

\begin{itemize}
\item Quicksilver~\cite{quicksilver} is an ECP proxy application that solves a simplified dynamic Monte Carlo particle transport problem, 
representing some elements of the workload of LLNL's Mercury~\cite{mercury} radiation transport code. 
\item PeleC~\cite{pelec} is an application using adaptive-mesh compressible hydrodynamics for reacting flows.
We studied its Premixed Flame (PMF) and Taylor-Green Vortex (TG) examples.
\item Nyx~\cite{nyx} is a massively parallel code computes compressible hydrodynamic equations on a grid with particles of dark matter.
We ran Nyx using up to 128 GPUs on Summit.
\end{itemize}

All three appliations were compiled with \texttt{-O3}.
In the remaining section, we describe insights based on the analysis result of HPCToolkit and its overhead on a single GPU and multi-GPU execution. 

\begin{figure*}[t]
\centering
\includegraphics[width=0.65\textwidth]{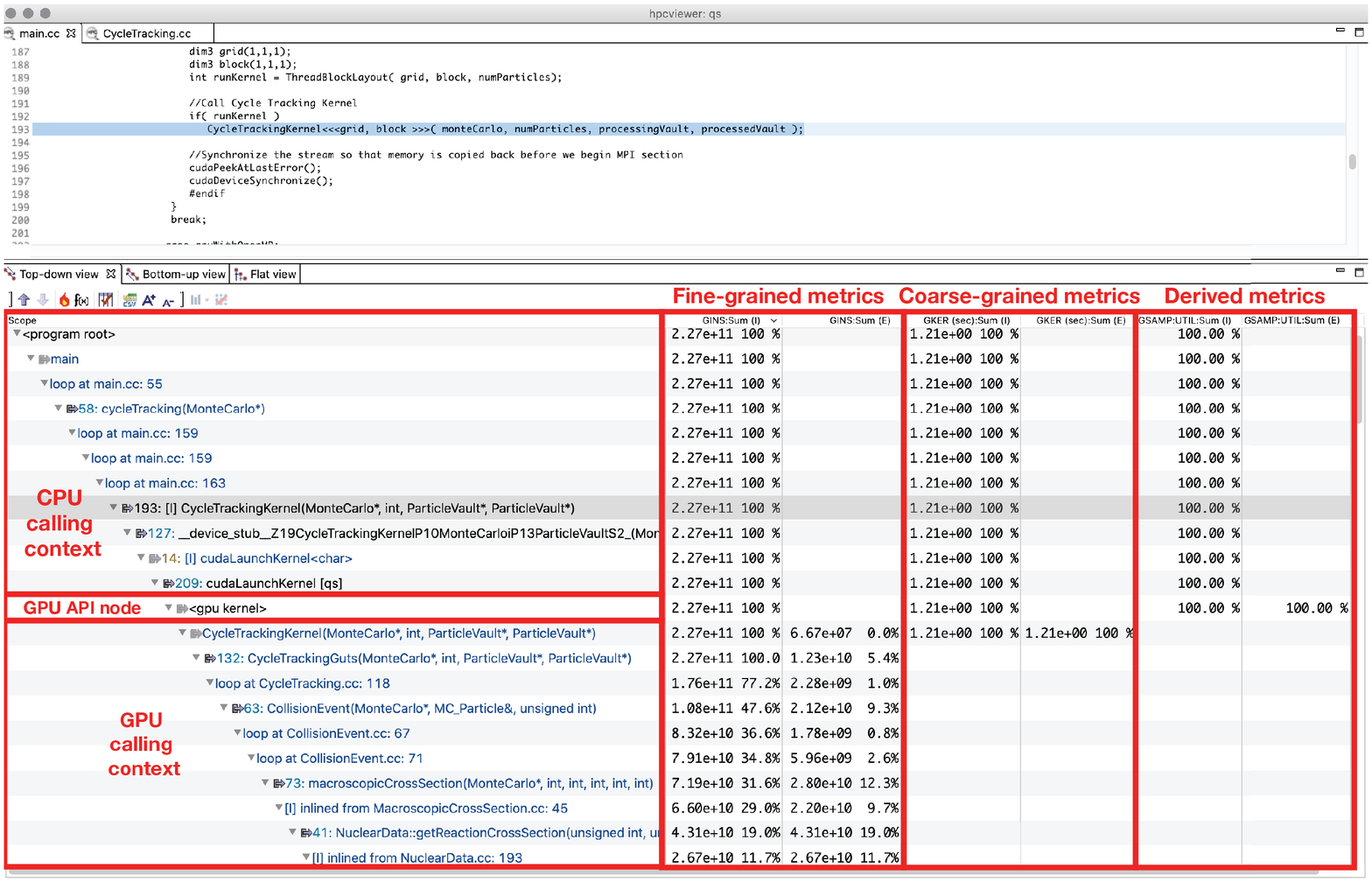}
\caption{Performance measurements of an optimized, GPU-accelerated
  execution of LLNL's Quicksilver proxy application on a Power9 and an NVIDIA GPU. Fine-grained
  measurements on the GPU were collected using PC sampling.}
\label{fig:quicksilver}
\end{figure*}

\subsection{Measurement Overhead}

On NVIDIA GPUs, HPCToolkit and nvprof introduce 2.24$\times$ and 2.20$\times$ profiling overhead for PeleC's TG example.
To trace Nyx on 128 ranks and 640 GPU streams, HPCToolkit and nvprof introduce 1.85$\times$ and 1.42$\times$ overhead.
It is worth noting that unlike nvprof, HPCToolkit collects CPU call stack information for every GPU API invocation.
On Intel GPUs, HPCToolkit's overhead for PeleC's TG application with
coarse-grained and fine-grained measurements is 1.81$\times$ and
2.23$\times$ respectively.  
AMD's ROCm software stack is evolving rapidly, and as a result, we
haven't assessed measurement overhead.
We have not yet invested much effort in analyzing and tuning
HPCToolkit's measurement overhead for any of the GPU-accelerated platforms.

\subsection{Analysis Overhead}
We compared the size of measurement data and analysis results for a GPU-accelerated version of Nyx using our original dense formats with that of our new sparse formats. 
The data in sparse formats is consistently much smaller than that using the original dense formats.
The size of the measurement results in the sparse format is 167.1 MB, which is 22$\times$ smaller than the dense format.
The size of the analysis results in sparse format is 153.3 MB, which is 3701$\times$ smaller than the dense format.

To measure analysis overhead, we used \textit{hpcprof-mpi} to aggregate and analyze measurements of a GPU-accelerated execution of LAMMPS~\cite{plimpton1995fast} on 167 nodes and 1002 GPUs.  Using thread-level parallelism and streaming aggregation, {\tt hpcprof-mpi} analyzed the 85GB of measurement data in 91 seconds on 48 nodes of Summit using 48 MPI ranks with 42 threads per rank (one thread per Power9 core). This is over 3.6$\times$ faster than using the original \textit{hpcprof-mpi}, which employs only inter-process parallelism, to analyze the data using the same resources.

\subsection{Quicksilver}

Figure~\ref{fig:quicksilver} shows the profile view for Quicksilver,
an ECP proxy application for LLNL's Mercury radiation transport
code. 
The bottom left pane shows a detailed heterogeneous calling context that spans both CPU and GPU.
The GPU calling context inside the {\tt CycleTrackingKernel} kernel is
reconstructed from flat PC samples using the algorithm described in Section~\ref{sec:gpucct}.
The bottom right pane shows two measured metrics (GPU instructions GPU kernel time), and a derived metric (GPU utilization).
Note that the code was compiled with \texttt{-O3}. Although many
device functions are inlined, others are not due to function size or register limitations.
With HPCToolkit's approximate GPU calling context tree reconstruction, 
an application developer can readily understand where GPU code is
costly and/or inefficient. (Understanding where a GPU code is
inefficient requires fine-grained stall metrics, which were collected
for Quicksilver although they are not shown in
Figure~\ref{fig:quicksilver}.)

\subsection{PeleC}
We studied PeleC's PMF problem on NVIDIA's GPUs only because there are not yet stable HIP or DPC++ implementations.
We ran PeleC's TG problem on the three platforms shown in
Table~\ref{tab:platforms} and compared the performance characteristics 
to understand relative performance of different GPUs and how efficiently the
software maps to each.

\subsubsection{NVIDIA GPU}
We profiled PeleC's PMF example with its default input using PC sampling. 
Using HPCToolkit's heterogeneous calling context, we were able to
identify the hot kernel quickly.
We noted that the GPU utilization for kernel
\texttt{pc\_expl\_reactions} was only 2.5\% on average, indicating low SM utilization.
By reducing the number of threads per block from 256 to 128, we increased the number of blocks for this kernel from 16 to 32 and improved its performance by 1.14$\times$.

\begin{figure}[htbp]
\centering
\includegraphics[width=0.45\textwidth]{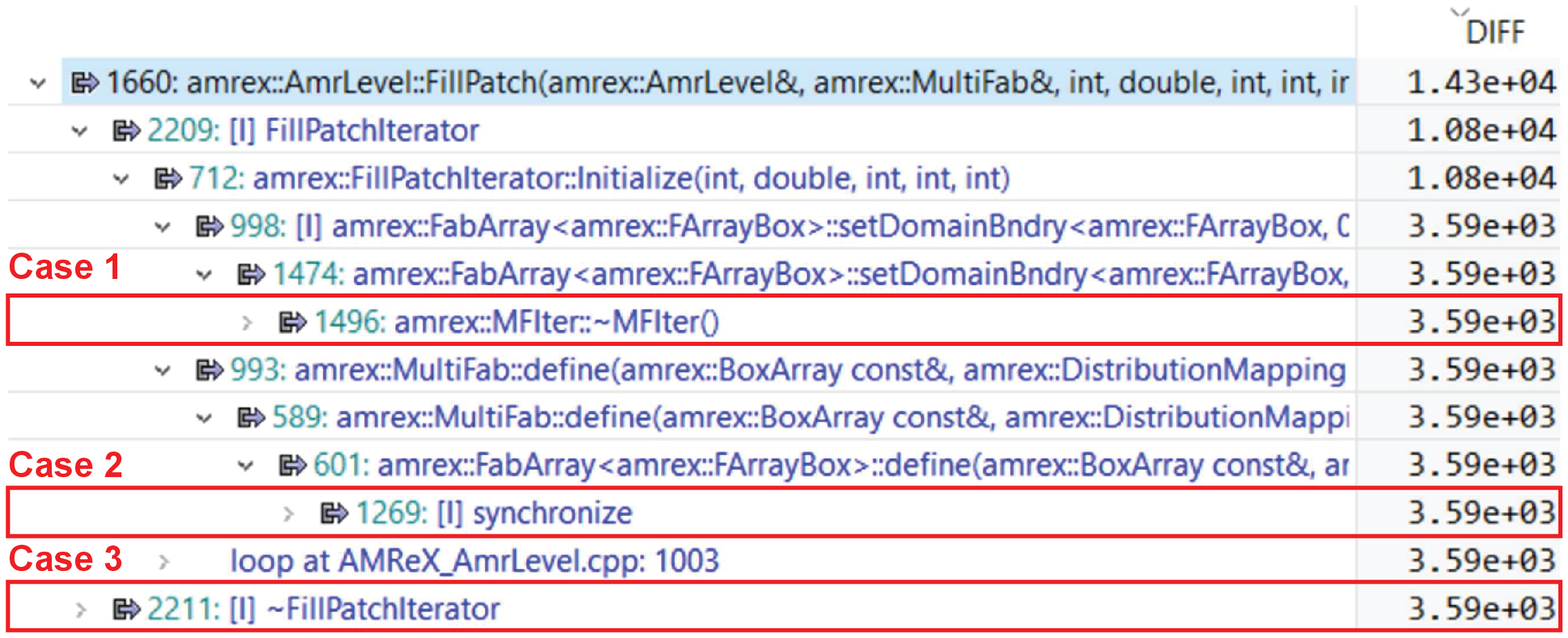}
\caption{An unnecessary synchronization calling context in PeleC}
\label{fig:synchronization}
\end{figure}

Next, we profiled PeleC's TG example with its default input.
\textit{hpcviewer} shows that the time spent on GPU synchronization is
close to GPU kernel execution time.
We found an unusual phenomenon: in some cases, the number of GPU synchronization API invocations exceeds the number of GPU kernel launches. 
We computed a derived metric: {\tt diff} $= sync\_count - kernel\_count$ in hpcviewer to find where synchronizations are unnecessary.
Figure~\ref{fig:synchronization} shows three calling contexts where {\tt diff} is high.
In the first calling context, no kernel launch occurs in the loop while there are synchronization invocations.
We found that an \texttt{MFIter} object is created for the loop, and the synchronization is called in the object's deconstructor to synchronize computations in the loop.
We optimized the code by not invoking synchronization if no computation is performed in the loop.
In the second calling context, the synchronization is needed only when there are multiple MPI ranks.
Thus, we can conditionally invoke synchronization by checking the number of MPI ranks.
In the third calling context, the synchronization is redundant because a copy function immediately before always synchronizes all GPU streams.
These three code changes reduced the number of synchronization invocations by 38\% and improved end-to-end performance by 1.05$\times$.

\subsubsection{Intel and AMD GPUs}
\label{sec:intel}

We compiled the DPC++ code of PeleC's TG example and ran it on an Intel Gen9 GPU using Intel's OpenCL backend.
We captured four coarse-grained metrics, including kernel execution time, memory transfer time, memory transfer bytes, and memory allocation bytes. 
For fine-grained measurement, HPCToolkit collected GPU instruction
counts using instrumentation added to GPU kernels using Intel's GT-Pin
library.

\begin{figure}[htbp]
\centering
\includegraphics[width=0.45\textwidth]{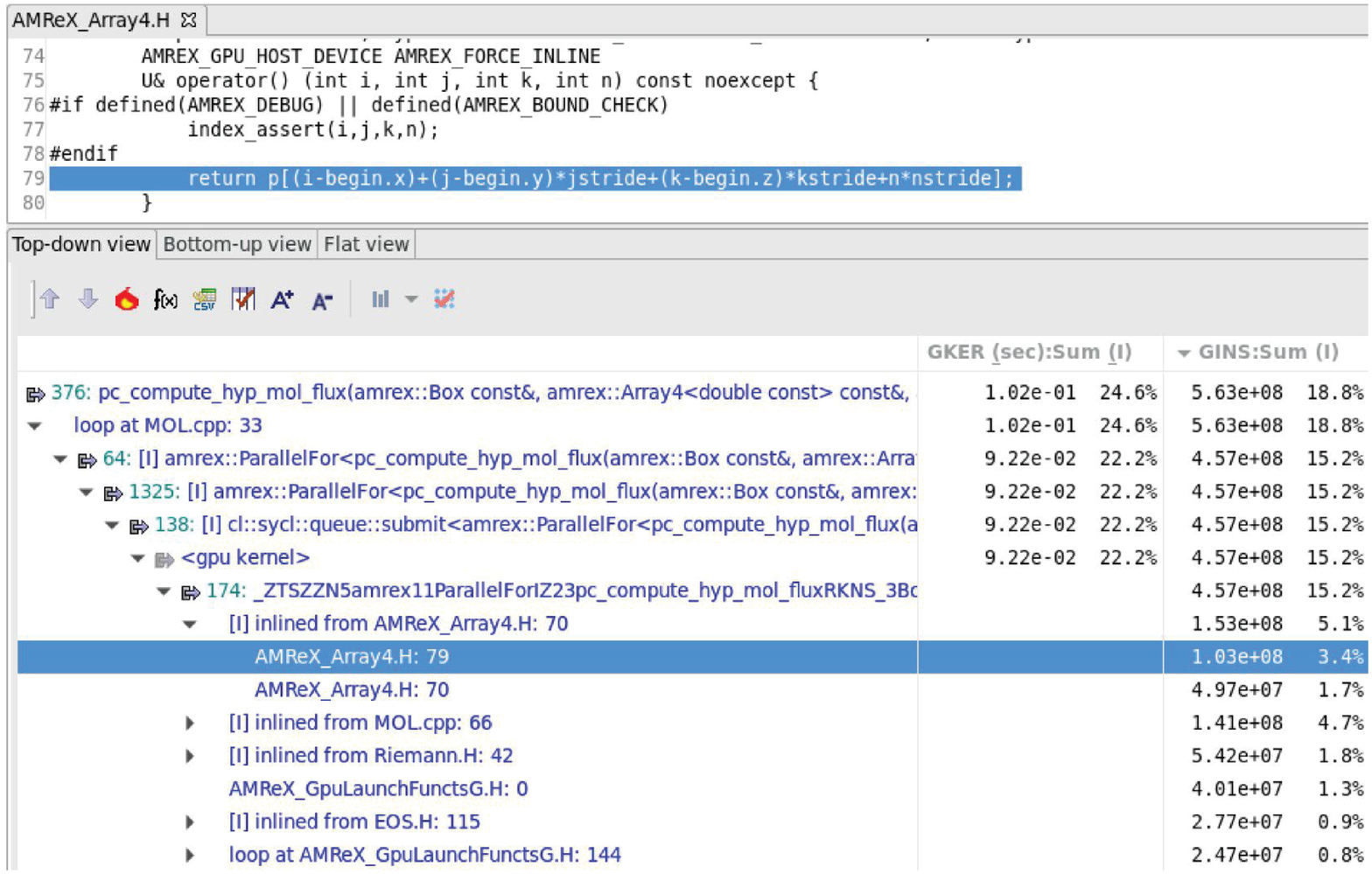}
\caption{Fine-grained measurements of PeleC's TG benchmark collected
  with binary instrumentation on an Intel Gen9 GPU. PeleC is implemented
  using DPC++ and executing GPU atop Intel's OpenCL runtime.}
\label{fig:intel_opencl_with_instrumentation}
\end{figure}

\begin{figure*}[t]
\centering
\includegraphics[width=.75\textwidth]{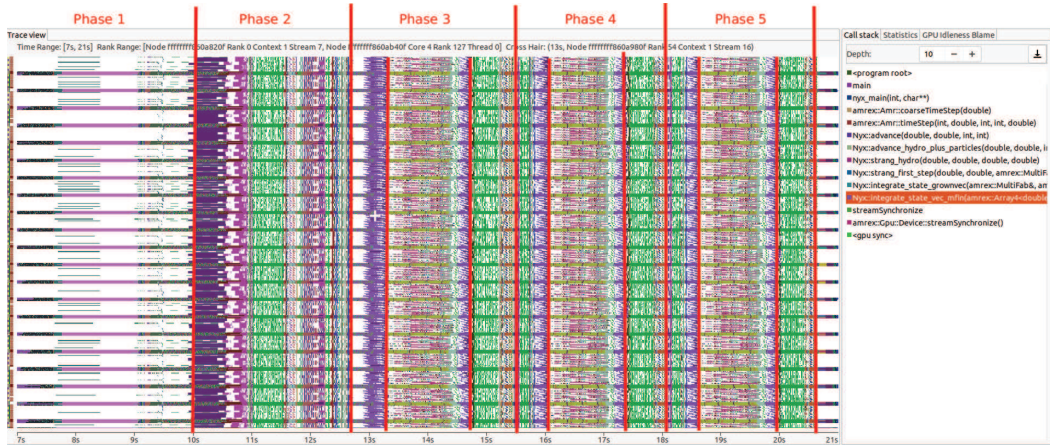}
\caption{Nyx' trace view running across 22 nodes using 640 GPU streams and 128 processes.}
\label{fig:hpcviewer_callstack}
\end{figure*}

The profile view in Figure~\ref{fig:intel_opencl_with_instrumentation}
enables one to quickly identify the most costly GPU kernels with
respect to GPU instruction count: \texttt{pc\_compute\_hyp\_mol\_flux}
and \texttt{pc\_compute\_diffusion\_flux}, which execute 37.6\% and 29.6\% of the GPU instructions respectively.
Measurement and attribution of instruction counts within each kernel revealed an interesting detail: roughly one third of the
instructions in each kernel are for array index calculation.
It is worth noting that the fraction of kernel execution time measured for this kernel while
collecting instruction level measurements may not be accurate
since kernel instrumentation inflates kernel execution time.

In addition, we noted $28.8E6$ bytes of GPU memory was allocated, but
no memory transfer were measured.
In this case, the memory transfers are implicit; thus, wrapping public
OpenCL APIs may miss observing internal memory transfers performed by a vendor's implementation.
To observe implicit memory transfers, we either need either vendors to
use only public APIs for data transfers or we need a better tool API.

In experiments with PeleC on AMD GPUs, HPCToolkit can collect
heterogeneous call stacks that attribute costs to GPU operations,
including kernel launches and data copies. Without support for
fine-grained measurements, we are unable to measure performance inside
GPU kernels.

\subsection{Nyx}

Figure~\ref{fig:hpcviewer_callstack} shows a trace view of Nyx
executing on Summit using 640 streams across 128 GPUs. The trace view shows that this execution consists of five phases.
By inspecting the call stacks in the trace view, we can determine what
the phases do, e.g., initializing particle and dark matter data,
performing hydrodynamic calculations, calculating gravity, etcetera. 

For each phase, we employed blame analysis to understand why GPUs are idle.
In the first phase, 58.01\% of idleness is caused by a call to
\texttt{cuCtxSynchronize}, which synchronizes all streams on the GPU.
Because only a single stream is used in this phase and a synchronized
memory copy always follows the call to \texttt{cuCtxSynchronize}, we
can safely remove the synchronization call and reduce the running time by 0.6s.

In the second phase, we easily identified that the major cause of
idleness is JIT compilation at runtime.
By providing the specific GPU architecture flag to the compiler and recompiling the program, we reduced the running time by 0.2s with this single optimization.

In the following three phases, we identified that idleness is caused
by calls to \texttt{MPI\_Waitall}, which suggests that there may be
opportunities for improving performance by optimizing communication.

The insights provided by blame shifting analysis reduced the GPU
running time of Nyx from 10.6s to 9.8s, achieving a 1.08$\times$
speedup with 640 GPU streams.
While the problems and improvements that we describe here are small, the important part is that HPCToolkit provides developers insights to identify even small problems. 

\section{Conclusions and Future Work}~\label{sec:conclusion}
Based on our experiences extending Rice University's 
HPCToolkit performance tools for GPU-accelerated platforms, 
we offer some observations about our perceptions of the 
hardware and software challenges faced by performance tools. 

\iparagraph{Measurement} To understand in detail the performance of complex GPU kernels,
hardware support for measuring GPU performance is essential. Today,
NVIDIA is the only vendor whose GPUs provide support for fine-grained
measurement using PC sampling. Without hardware support for fine-grained
performance measurement, application developers must use the ``guess
and check'' strategy for performance optimization, which significantly
complicates the process of analysis and optimization of large-scale
applications.

Today, NVIDIA's PC sampling measurement serializes execution of GPU kernels to simplify performance
measurement. This approach both slows the execution of applications
when under observation by performance tools and distorts the
performance of applications available for tools to observe.

Another problem tools face is that not all GPU activities are readily
observable. As described in~\ref{sec:intel}, there is no portable
mechanism a tool can use to monitor implicit buffer transfers in
OpenCL. In addition, NVIDIA purposefully avoids having CUPTI monitor
implicit synchronization invoked by primitives such as {\tt
  cudaMemcpy} as well as driver calls and other operations called by
NVIDIA libraries such as {\tt cuBlas}~\cite{welton2019diogenes}.
The inability to monitor all data transfers and
synchronization gives tools an incomplete picture of an application's
behavior. For tools to be maximally effective, all GPU operations
must be observable. 

\iparagraph{Attribution} To help application developers understand the performance of optimized 
GPU kernels that are generated
from or employ inlined templates or functions, tools need compilers
to generate high-quality DWARF information that describes the provenance of 
every machine instruction in each GPU kernel. While line mappings generated
by today's compilers relate most machine instructions back to an associated
source line, they often fail to relate machine instructions back to
any inlined call chain that caused it to be included.  
In our experience, even the latest compilers for 
AMD, Intel, or NVIDIA GPUs do not generate sufficiently precise DWARF for 
attributing the performance of each GPU instruction to its full calling 
context in the presence of inlined functions and templates.

\iparagraph{Analysis} Measuring and attributing the performance of GPU kernels with hardware support
for fine-grained measurement is
only the first step towards understanding the performance of complex applications.
Within GPU kernels, understanding the causes of performance problems and opportunities for improvement
requires understanding the interplay between a wide range of factors, including block-level parallelism,    
thread-level parallelism within blocks, data sizes, and alignment and how it maps into the memory hierarchy,
details of generated code (e.g., resource consumption, use of type conversions, use of special 
functional units, and instruction sequencing), and reasons for stalls that result in uncovered latency.

While our work developing a framework for scalable analysis of
perforamnce data for extreme-scale executions is an important building block,
it is clear that our tool could use further tuning to better exploit the capabilities of parallel file systems found on supercomputers. 
Also, it would be beneficial if we could exploit accelerators in our
analysis.

\iparagraph{Presentation} The most common performance problem we have observed in 
emerging ECP applications and libraries is that GPUs are idle too much of the time. 
Understanding the root causes of idleness in large-scale applications composed of many 
layers of software is difficult for application developers using existing user interfaces.
Our initial prototype support for automated analysis of GPU traces offers some help in this regard, but our early experiences have shown that
attributing performance at the proper level to generate insight will require new design insights and implementation.

\section{Acknowledgements}
\sloppy
This research was supported in part by the Exascale Computing Project 
(17-SC-20-SC)---a collaborative effort of the U.S. Department of Energy Office of Science and the National Nuclear Security Administration, 
Argonne National Laboratory (DE-AC02-06CH11357), DOE Tri-labs (LLNL Subcontract B639429),
and Advanced Micro Devices, Inc..

\sloppy
This research used resources of the Oak Ridge Leadership Computing Facility at the Oak Ridge National Laboratory, which is supported by the Office of Science of the U.S. Department of Energy under Contract No. DE-AC05-00OR22725, and Argonne Leadership Computing Facility, 
which is a DOE Office of Science User Facility supported under Contract DE-AC02-06CH11357.


\end{document}